\documentclass[aps,prb,twocolumn]{revtex4}

\usepackage{graphicx}

\begin{document}

\setlength{\pdfpagewidth}{8.5in}
\setlength{\pdfpageheight}{11in}

\title{Critical Waves and the Length Problem of Biology}

\author{Robert B. Laughlin}

\affiliation{Department of Physics, Stanford University,
Stanford, CA 94305}

\begin{abstract}
It is pointed out that the mystery of how biological systems measure 
their lengths vanishes away if one premises that they have discovered a 
way to generate linear waves analogous to compressional sound. These can 
be used to detect length at either large or small scales using echo 
timing and fringe counting. It is shown that suitable linear chemical 
potential waves can, in fact, be manufactured by tuning to criticality 
conventional reaction-diffusion with a small number substances.  Min 
oscillations in {\it E. coli} are cited as precedent resonant length 
measurement using chemical potential waves analogous to laser detection. 
Mitotic structures in eucaryotes are identified as candidates for such 
an effect at higher frequency. The engineering principle is shown to be 
very general and functionally the same as that used by hearing organs.
\end{abstract}

\homepage[R. B. Laughlin: ]{http://large.stanford.edu}

\date{March 1, 2015}

\maketitle

It is not known how living things measure their lengths. This is true 
notwithstanding the immense progress made over the past 30 years in 
understanding morphogen gradients in embryogenesis. 
\cite{nusslein,stathopoulos,sick,economou,sheth,muller}.  The problem is 
captured nicely by the confusion over regulation of the bicoid profile 
in {\em Drosophila} \cite{gregor,spirov,lipshitz,grimm,cheung}, but it 
is also reflected in the notorious instability, hysteresis, and lack of 
scalability of traditional static reaction-diffusion 
\cite{koch,pearson}. No one knows why cells are the size they are 
\cite{marshall}, why plants and animals are the size they are 
\cite{nijhout}, how organs grow maintaining their proportions 
\cite{stanger}, and how some animal bodies regenerate lost limbs 
\cite{king}. On the matter of length determination, per se, very little 
progress has been made beyond Thompson's 1917 treatise on biological 
form \cite{thompson}.

Length has a special place in biology by virtue of being a primitive 
quantity with units. It is not possible for living things to size 
themselves properly without having developed the skill of measuring 
these quantities as numbers and relating these numbers to each other 
mathematically. They require meter sticks to do this.  They must 
fabricate these meter sticks using diffusion and motors, since they are 
the only biochemical elements that involve length. The relationships of 
these meter sticks to each other and to the lengths they measure must be 
precise and described by equations. This is because precise mathematical 
relationships among lengths are what size and shape are.

In this paper I point out that the difficulty of accounting for length 
relationships of parts of organisms with equations disappears instantly 
if the organism is premised to have discovered a way to emulate 
elementary physical law. In particular, one simple invention is 
sufficient to facilitate the measurement and construction of body plans 
of any shape and size one might wish in a way that is both plastic and 
scalable: the conversion of diffusive motion into {\it linear} waves 
using engines. The concept is general because all motion in the presence 
of disorder, including motion of cytoskeletal components, becomes 
diffusive at long time and length scales by virtue of evolving into a 
random walk.  But if engines can transform this random walking into 
propagation with stable direction and speed, then signals can be beamed, 
like a flashlight, reflected from boundaries, and trapped. Once this 
happens, the organism can measure lengths the same way human engineers 
do, by echo timing or by fringe counting and resonance. Body designs 
based on this strategy are inherently plastic because fixing the speed 
enables lengths to be laid out or detected by means of clock tick 
intervals, which are easy to change.

\begin{figure}
\centerline{\includegraphics[width=.45\textwidth]{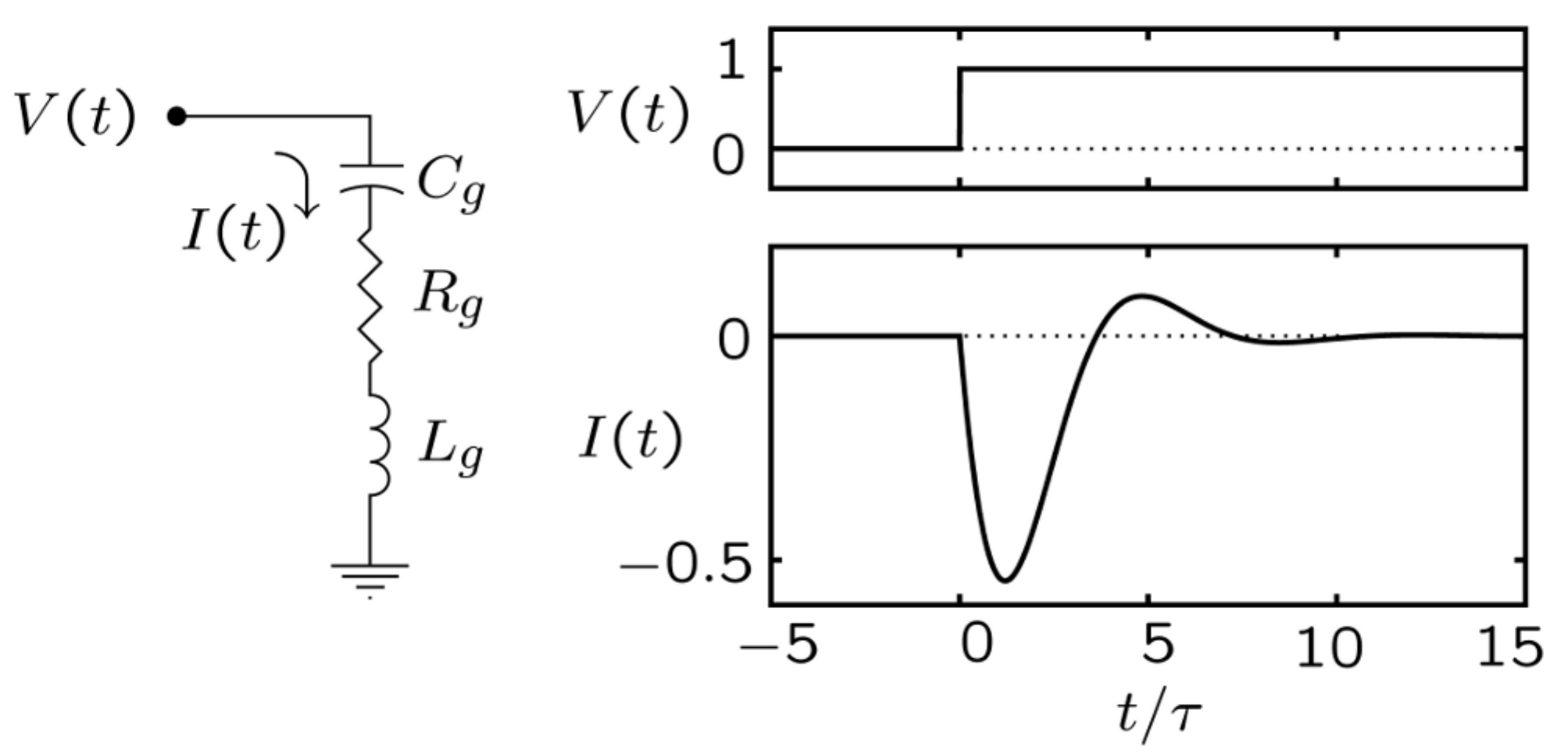}} 
\caption{Chemical amplifier described in electrical terms. The 
circuit components $R_g$, $L_g$ and $C_g$ all have negative values. On 
the right is shown the impulse response described by Eq. (\ref{impulse}) 
for the special case of $R_g C_g = L_g / R_g = \tau$.  The voltage step 
height $V_0$ is set to 1. The current response, plotted as a multiple of 
$V_0/|R_g|$, is negative and has the same shape as a neuron action 
potential.}
\label{f1} 
\end{figure}

Although it is not widely known, biological systems can easily 
manufacture such waves using elementary reaction-diffusion chemistry 
similar to that at work in neuron action potentials. \cite{turing} The 
key is tuning the chemical reactions to the edge of an instability, an 
effect known in the cochlear amplifier literature as criticality. 
\cite{gold,camalet,hudspeth} As I shall show, this trick is so easy to 
implement technically that it is hard to imagine how Nature would not 
have exploited its tremendous engineering advantages in the struggle for 
survival. This obligates us to take seriously even very small hints that 
Nature did, in fact, discover how to do it long ago. The simple 
explanation for why we have found only sparse empirical evidence for 
such waves so far is that chemical potential waves are difficult to 
detect with existing laboratory techniques. The larger idea implicit in 
this proposed resolution of the length problem is that biological 
systems cannot conduct engineering without rules any more than we humans 
can, so they invented some in the ancient past, and the ones that worked 
best turned out to be same ones we humans discovered later using reason.
 
\section{Chemical Potential Waves}

The simplest chemical length mensuration apparatus involves chemical 
potential waves solely. Apparati with other components, such as 
mechanical motors, are allowed also, but all of them necessarily have a 
chemical potential component by virtue of how they work.

Despite being difficult to detect, chemical potential waves are known to 
be pervasive in biology. The most familiar case is the neuron action 
potential, the electrical aspects of which make it easy to detect by 
primitive means, even though it is fundamentally a reaction-diffusion 
wave \cite{scott}. But there are also non-electrical varieties: cAMP 
waves in slime molds, which direct the colony's organization into 
fruiting bodies \cite{goldbeter}; calcium waves, directly implicated in 
oogengenesis \cite{jaffe,jaffe1}, developmental patterning 
\cite{whitaker}, brain function \cite{weissman,scemes,kuga}, and cell 
signaling in animals \cite{junkin} and plants \cite{choi}; and MinDE 
waves in {\em E. coli}, perhaps the most important of all because they 
are involved in a bacterial length decision \cite{lenz}. All of these 
non-electrical versions require highly advanced technologies to see, and 
also required a bit of luck to find, so there is good reason to suspect 
that more exist and simply have not been detected yet.

The simplest way chemical reactions can manufacture waves is through 
stable two-terminal amplification, the fundamental basis of laser 
operation \cite{siegman}.  The observation that amplification is 
involved is important, for while all amplifiers exploit nonlinearities 
to work, there is nothing inherently nonlinear about what they do.  All 
amplifiers become nonlinear when they are pushed to deliver large 
powers. It is thus not surprising that the wave signals easiest to 
observe in biology are often nonlinear.  But amplification in the linear 
regime is known to occur as well, notably in hearing organs 
\cite{gale,ashmore,manley,warren,mhatre,mora}.

\section{Amplifiers: Stability and Causality}

Two-terminal amplifiers are easiest to explain by electrical analogy.  
Consider the circuit shown in Fig. \ref{f1}. It is a conventional linear 
resonator, such as one might find in any radio, except that the 
components all have negative values. Its active component, the negative 
resistor, causes electric current to flow in a direction opposite to the 
way it normally would when voltage is applied. Such reversed flow is 
implicit in all Na$^+$-K$^+$ neural models, including the original one 
of Hodgkin and Huxley \cite{hodgkin}, but it is quite explicit in those 
based on tunnel diodes \cite{fitzhugh}. The all-important $C_g < 0$ 
causes induced current to stop flowing after a time $R_g C_g$.  Na$^+$ 
channels achieve this cessation by plugging themselves after a time 
delay with a molecular stopper \cite{goldin}. The $L_g < 0$ mainly 
causes a finite turn-on time $L_g/R_g$, but it also adjusts the 
circuit's after-bounce, so it corresponds to the K$^+$ channel of a 
neuron. Thus Fig. \ref{f1} is simply a linearized version of the 
Hodgkin-Huxley equations.

\begin{figure}
\centerline{\includegraphics[width=.45\textwidth]{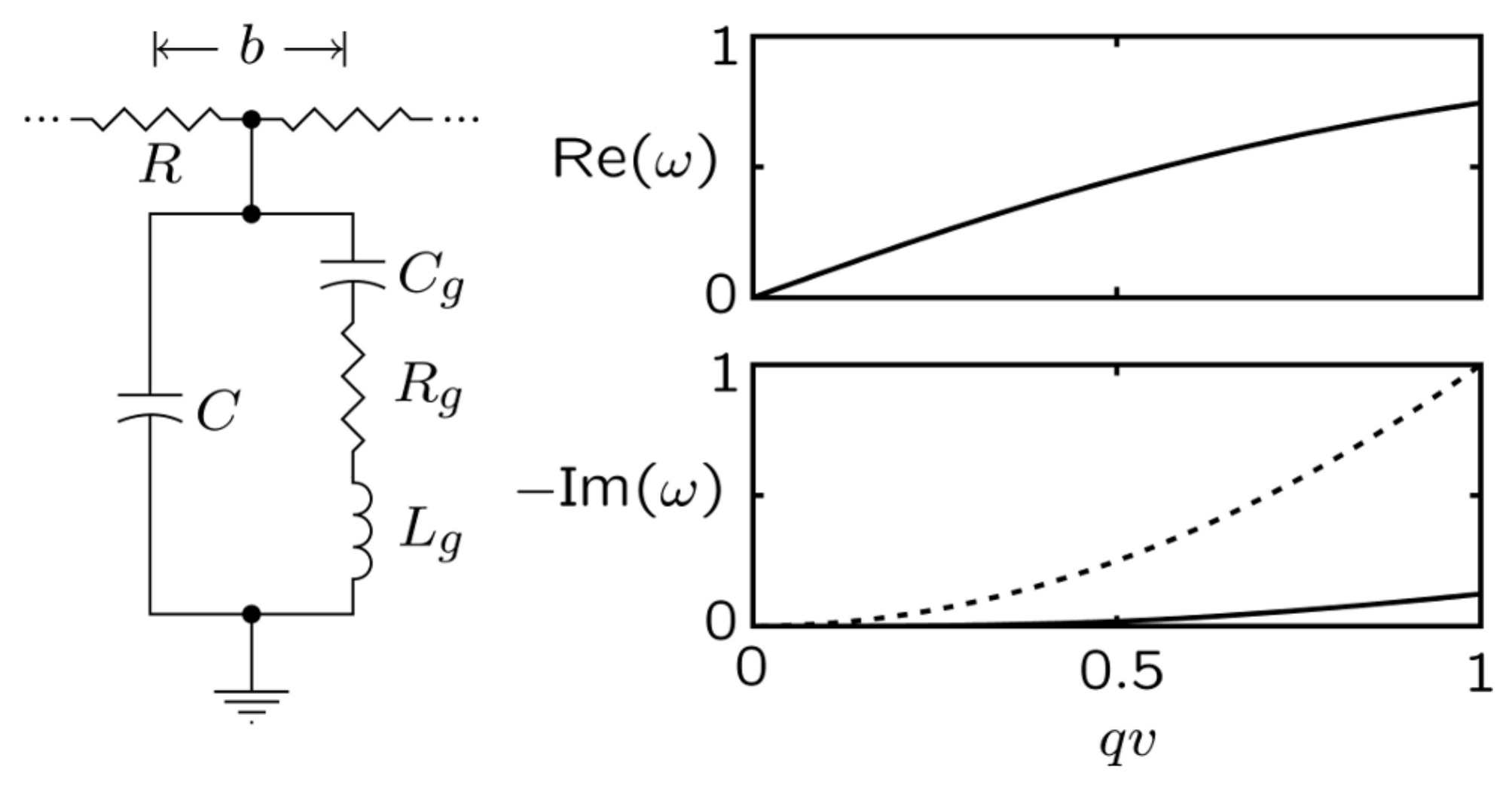}} 
\caption{Left: Illustration of a diffusive transmission line with a 
series of amplifiers like those in Fig. \ref{f1} placed across it. R and C 
represent the resistance and capacitance per unit length before the 
amplifier is added. The repeat distance is $b$. Right: Plot of the 
solution of Eq. (\ref{dispersion}) for the special case of $C + C_g = 0$ 
and $R_g C_g = L_g/R_g = \tau$.  Both $\omega$ and $qv$ are expressed as 
multiples of $1/\tau$. The complex numbers $\omega$ and its negative 
complex conjugate $-\omega^*$ are two of the three roots, the third 
being pure imaginary and off the top of the graph. The dashed line is 
the solution when the amplifiers are turned off ($C_g \rightarrow 0$).}
\label{f2} 
\end{figure}

It is important that both $C_g < 0 $ and $L_g < 0$ are dynamical 
creations of the amplifier itself, not additional postulates. It is 
physically impossible to make a stable amplifier without also creating, 
as a side effect, negative reaction. This effect is seen in lasers as a 
reversal of the dielectric function whenever the laser gain medium 
becomes amplifying \cite{desurvire}; but the deeper reason has nothing 
to do with quantum mechanics or population inversions.  It is causality 
\cite{toll}. The current induced by a stimulating voltage can appear 
only after the stimulus is applied, never before.  The current induced 
by the step voltage shown in Fig. \ref{f1} is

\begin{equation}
I(t) = \frac{V_0}{2\pi} \int_{-\infty}^\infty \frac{C_g
e^{-i \omega t}}{1 - i \omega R_g C_g - \omega^2 L_g C_g}
\; d\omega
\label{impulse}
\end{equation}

\noindent
It is properly causal provided that poles of the response kernel lie
in the lower half of the complex plane. This requires both
$C_g$ and $L_g$ to be negative if $R_g$ is.

The physical principles operating in Fig. \ref{f1} apply to all 
amplifiers, not just electrical ones. The denominator in Eq. 
(\ref{impulse}) may be seen to be a Taylor expansion in $\omega$ 
truncated to second order. Such a truncation is always valid at long 
times, and it is equivalent to stating that the system has only two 
poles in the complex plane and thus has only two important mechanical 
degrees of freedom.  Those things are therefore not model assumptions at 
all but generic features of amplifier response at long times. In the 
case of Fig. \ref{f1}, the degrees of freedom are charge and current, 
but in general they could be anything.

\section{Turing Critical Wavefunction}

Linear waves are produced when one places a series of such amplifiers 
across a diffusive transmission line, as shown in Fig. \ref{f2}. This is 
aptly analogous to placing Na$^+$-K$^+$ amplifiers across the membrane 
of an axon. Substituting $V_j = V_0 \, \exp[i(qbj - \omega t)]$, where 
$b$ is the repeat length, for the voltage on the jth site, we obtain the 
dispersion relation

\begin{equation}
\frac{2}{R} [ 1 - \cos(qb)] - i \omega C
- \frac{i \omega C_g}{1 - i \omega R_g C_g - \omega^2
L_g C_g} = 0
\label{chain} 
\end{equation}

\noindent
When the amplifiers are turned off ($C_g \rightarrow 0$), this becomes 
the diffusion equation

\begin{equation}
D q^2 - i \omega = 0
\; \; \; \; \; \; \; \; \; \; \; \; \; \; \; 
(D = \frac{b^2}{RC})
\end{equation}

\noindent
in the limit of small $q$. But when the amplifiers are turned on, and 
also adjusted so that $C + C_g = 0$, Eq. (\ref{chain}) becomes

\begin{equation}
(vq)^2- \frac{\omega^2 ( 1 - i \omega \tau)}
{1 - i \omega \tau - \omega^2 \tau^2} = 0
\; \; \; \; \; \; \; \; 
(v = \sqrt{D / \tau})
\label{dispersion}
\end{equation}

\noindent
This is functionally equivalent to the wave equation

\begin{equation}
\frac{\partial^2 \psi}{\partial x^2} = \frac{1}{v^2} 
\frac{\partial^2 \psi}{\partial t^2}
\; \; \; \; \; \; \; \; \; \; \; \;
(v = \sqrt{D / \tau})
\label{wave}
\end{equation}

\noindent
where $\psi$ is any one of the dynamical variables, in the regime $qv < 
0.5/\tau$. The expression for the velocity $v$ is the Luther equation 
\cite{showalter}. The full solution Eq. (\ref{dispersion}) is plotted in 
Fig. \ref{f2}.

This wave equation is equivalent to the Turing reaction-diffusion 
equations

\begin{eqnarray}
\frac{\partial X}{\partial t} &=& \frac{1}{\tau} Z
+ D \frac{\partial^2 X}{\partial x^2}
\nonumber\\
\frac{\partial Y}{\partial t} &=& \frac{1}{\tau} Z 
\nonumber\\
\frac{\partial Z}{\partial t} & =& \frac{1}{\tau}
(X - Y - Z) 
\label{turingeq}
\end{eqnarray}

\noindent 
It could thus easily be achieved with chemical reactions among 
three substances. It is also easily generalized to three dimensions.  In 
fact, the principle behind Eq. (\ref{dispersion}) is so general that it 
applies to any conservative diffusive phenomenon, chemical, electrical 
or mechanical, at any length or time scale, regardless of details. For 
this reason, it is a competitive candidate for how living things might 
measure their lengths {\it generally}.

Stabilization of the wave velocity in these systems is achieved through 
two fine-tunings.  One is equality of the capacitive and inductive 
times. This is a matter of amplifier design and is achieved in the case 
of neurons by having the right mix of Na$^+$ and K$^+$ channels. The 
other is cancellation of the capacitances, a result achieved in practice 
by slowly increasing the number of amplifiers in the membrane until the 
system begins to oscillate a little.  Such oscillations are routinely 
observed emanating from the ear \cite{kemp}. A similar phenomenon has 
been reported at the surface membranes of yeast \cite{pelling}.

\begin{figure} 
\centerline{\includegraphics[width=.45\textwidth]{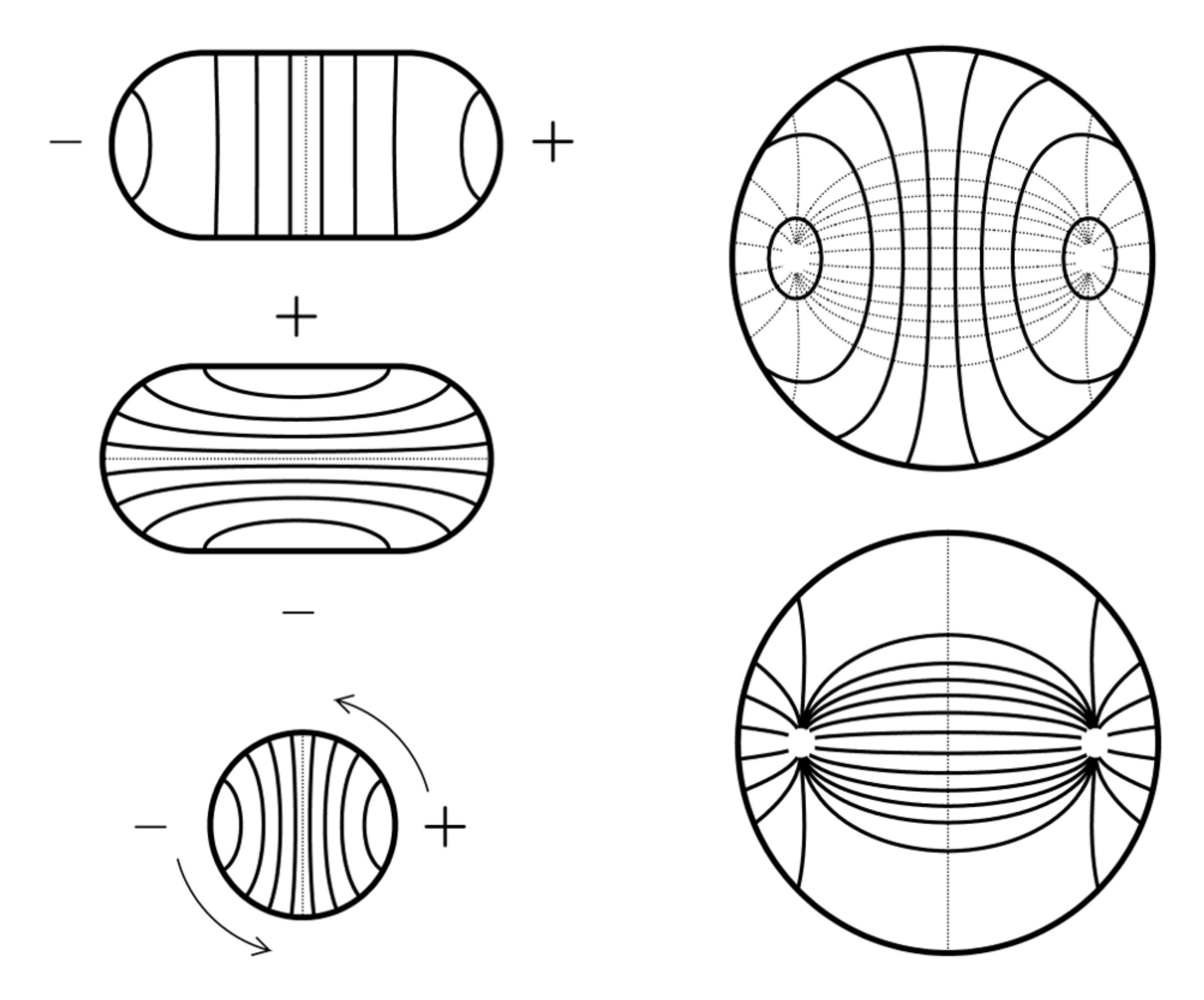}} 
\caption{Left: Solutions of Eq. (\ref{helmholtz}) with Neumann boundary 
conditions for a pill-shaped cavity, as appropriate for a bacterium.  
Hemispherical end caps of radius $a$ are attached to a cylinder of 
length $2.2 \, a$. The top shows a contour plot of the lowest 
eigenfunction, which corresponds to an axially symmetric pole-to-pole 
sloshing. It occurs at $\omega_n = 0.88 v / a$. Below that are contour 
plots of first longitudinally symmetric mode, which corresponds to a 
steady migration around the perimeter along a path perpendicular to the 
axis. It occurs at $\omega_n = 1.91 v/a$.  This mode is degenerate with 
a mirror-reflected one that rotates in the opposite direction. Right: 
Solution of Eq. (\ref{helmholtz}) for a spherical cavity with mixed 
boundary conditions ($\hat{{\bf n}} \cdot {\bf \nabla} \phi_n = - 0.51 
\phi_n$). The top shows a contour plot of the lowest eigenfunction.  The 
dotted lines, reproduced as solid lines below, are trajectories 
perpendicular to the contour lines.}
\label{f3} 
\end{figure}

\section{Length Measurement Using Cavity Resonance}

The simplest strategy for measuring length with manufactured waves is 
detecting mode resonances in a cavity.  This is illustrated in the case 
of 3-dimensional waves in Fig. \ref{f3}. The allowed oscillations of a 
cavity are found by substituting a harmonic solution $\psi({\bf r},t)= 
\phi_n({\bf r}) \, \exp(- i \omega_n t)$ into Eq. (\ref{wave}) and 
solving the Helmholtz equation

\begin{equation}
\nabla^2 \phi_n + (\omega_n / v)^2 \, \phi_n = 0
\label{helmholtz}
\end{equation}

\noindent 
with Neumann boundary conditions ($\hat{{\bf n}} \cdot {\bf \nabla} 
\phi_n = 0$), as appropriate for a substance that is conserved and 
cannot flow in or out through the walls. Solutions exist only for 
certain discrete eigenfrequencies $\omega_n$. The values of these and 
the spatial behavior of their corresponding eigenfunctions $\phi_n$ 
sense the size and shape of the cavity. The eigenfunction $\phi_n$ 
corresponding to the lowest of eigenfrequency shown in in Fig. \ref{f3} 
describes to the observed MinDE wave motion in {\it E. coli}, although 
details differ.

\begin{figure} 
\centerline{\includegraphics[width=.45\textwidth]{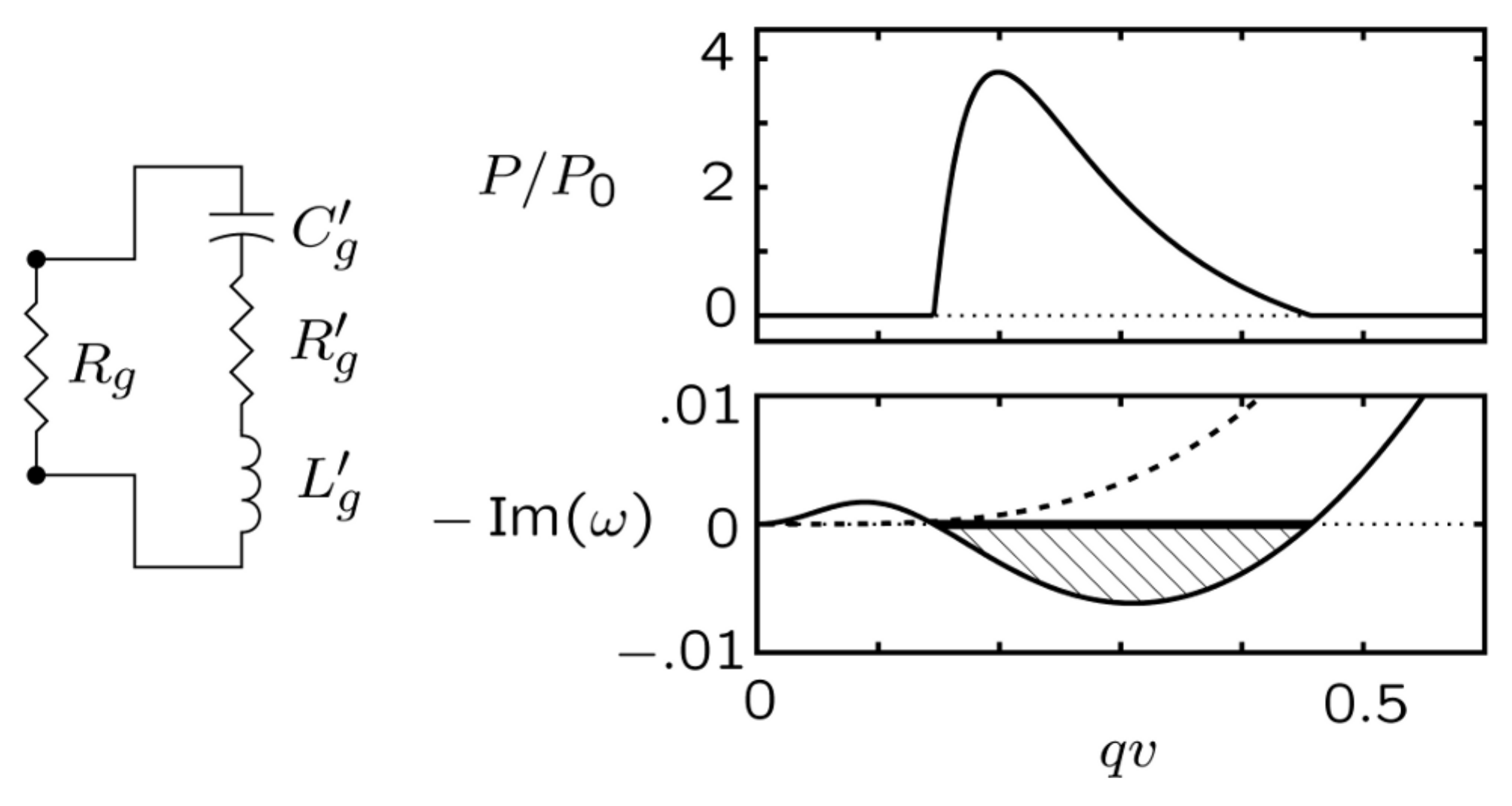}}
\caption{Left: Illustration of the modification of $R_g$ required to 
introduce a gain peak into the transmission line of Fig. \ref{f2}, as 
described by Eq. (\ref{gainpeak}). Lower Right: Correction to dispersion 
relation of Fig. \ref{f2} resulting from the values $\omega_0 = 
0.2/\tau$, $f_0 = 0.1$, and $\tau_0 = 10 \tau$.  Only the imaginary part 
of $\omega$ is shown because the correction to the real part is 
negligible. Hatching indicates the region of oscillation. Here 
saturation, described by Eq. (\ref{saturation}), pushes Im($\omega$) to 
zero. The dashed line shows the $C_g' \rightarrow 0$ behavior. Both 
$\omega$ and $q v$ are expressed as multiples $1/\tau$. Upper Right: 
Power produced at saturation, per Eq. (\ref{saturation}).}
\label{f4} 
\end{figure}

To actually excite eigenmodes of a cavity it is necessary provide the 
amplifying medium with a gain peak.  This is achieved most simply in the 
case of the example of Fig. \ref{f2} by attaching a second resonant 
circuit, as shown in Fig. \ref{f4}. This effectively makes the 
amplifying resistor slightly frequency dependent, per

\begin{equation}
\frac{1}{R_g} \rightarrow 
\frac{1}{R_g} \biggl\{ 1 + f_0 \biggl[
\frac{- i \omega \tau_0}{1 - i \omega \tau_0 - (\omega/\omega_0)^2}
\biggr] \biggr\}
\label{gainpeak}
\end{equation}

\noindent 
where $\tau_0 = R_g' C_g'$, $\omega_0 = (L_g' C_g')^{-1/2}$ and $f_0 = 
R_g/R_g'$. Modifying $R_g$ in this way is physically equivalent to 
humming a tone in a closed room: The tone is $\omega_0$, the loudness is 
$f_0$, and the time between successive breaths is $\tau_0$.  The 
corresponding reaction-diffusion equations are

\begin{eqnarray}
\frac{d X}{d t} &=&  D \frac{\partial^2  X}{\partial x^2} 
+ \frac{1}{\tau} Z\nonumber\\
\frac{d Y}{d t} &=&  \frac{1}{\tau} Z\nonumber\\
\frac{d Z}{d t} &=& \frac{1}{\tau} (X - Y - Z + f_0 Z')\nonumber\\
\frac{d Y'}{d t} &=& \frac{1}{\tau_0} Z'\nonumber\\
\frac{d Z'}{dt} &=& \omega_0^2 \tau_0 \bigl[Z - Y'
- (1 + f_0) Z' \big]
\label{bigturingeq}
\end{eqnarray}

When modified in this way the system becomes a textbook laser 
oscillator. As shown in Fig. \ref{f4}, cavity modes with frequencies 
$\omega_n$ in the region of net gain grow exponentially and saturate the 
amplifier, meaning they eat up all the power available.  In biological 
terms we would say that a nonlinearity chokes off the exponential growth 
and causes it to plateau.  Laser saturation reduces the gain according 
to the approximate formula

\begin{equation}
f_0 \longrightarrow f = \frac{f_0}{1 + P/P_0}
\label{saturation}
\end{equation}

\noindent
where $P$ is the power delivered and $P_0$ is a parameter characteristic 
of the medium.  $P$ continues to increase until Im($\omega$) becomes 
zero, as shown in Fig. \ref{f4}, for the mode with the highest native 
gain. This implies that all the other modes in the saturated state have 
negative gain and die away.  This winner-take-all competition for the 
available energy causes the system to oscillate in one mode rather than 
many.

The frequency and magnitude of the saturated oscillation both measure 
the cavity length.  The coarse-grained measurement is the discrete 
frequency jumping that occurs as one mode after another becomes dominant 
as the cavity is lengthened.  The power $P$ of the victorious mode is 
also modified by the length adjustment through the medium's gain 
profile, as shown in Fig. \ref{f4}. The saturated power thus provides a 
fine-tuning measurement of length.

\section{Precedent in Bacteria}

Gain, oscillation and saturation have all been observed experimentally 
for MinDE oscillations in {\it E. coli}. \cite{raskin}. Fluorescence 
tagging experiments have revealed that MinD molecules flock together 
from one end of the bacterium to the other with a round-trip travel time 
of about 40 seconds. Disabling expression of FtsZ, a protein required 
for septation and division, causes the bacterium to grow very long and 
exhibit a preferred MinD wavelength of approximately 10 $\mu$m, or about 
twice the length at which it normally divides \cite{raskin}. Both 
standing waves and traveling waves are observed in these long mutants, 
depending on circumstances. \cite{raskin,meacci-thesis} The system can 
also flip unstably between the two when the boundary conditions are 
changed.  When the bacterium divides, the oscillation bifurcates 
unstably and then settles down with a higher frequency, just as a laser 
would \cite{juarez}. The increase is measured to be a factor of 1.5, 
whereas a factor of 2 would be expected of perfectly linear waves.

There is no direct evidence that the bacteria oscillate for the purpose 
of measuring their lengths absolutely, nor is there any direct evidence 
that oscillations are universally present in all bacteria. One knows for 
certain only that {\it E. coli} use Min oscillations are part of the 
machinery for determining their midpoint for division, and that a MinD 
homolog in {\it B. subtilis} has been observed to form static patterns 
that do not oscillate. \cite{marston}

However, circumstantial evidence is abundant. The existence of Min 
oscillations clearly demonstrates that chemical reactions actually 
present in a bacterium have the ability to manufacture chemical waves 
and trap them. Nature presumably created this machinery for some 
purpose.  Min oscillations are known to involve only a small number of 
substances.  The exact number is controversial, but the reactions are 
typically modeled with four or five, the same as in Eqs. 
(\ref{bigturingeq}). \cite{howard,meinhardt5,huang,bonny,halatek}. 
Existing experiments are not sufficiently detailed to distinguish among 
these models, but elementary reaction-diffusion is central to all of 
them, and all become similar to Eqs. (\ref{bigturingeq}) when 
linearized. The fundamental simplicity of the reactions was indicated 
early on by identification of MinCDE operon damage as the cause of the 
minicell mutation in {\it E. coli} \cite{adler,deboer}, but it is now 
corroborated by experiments {\it in vitro} showing both oscillations and 
spatial waves occurring in system containing only MinD, MinE, a lipid 
membrane, and ATP \cite{loose,ivanov,schweizer}. The frequencies and 
lengths observed in these experiments do not agree with those observed 
in real bacteria, but this is not surprising given how finely tuned a 
reaction-diffusion system must be to measure lengths accurately.  They 
are like a watch with a corrupted regulator: It still ticks, but it does 
not keep time.

It is not important that the Min amplifier machinery resides in or near 
the cell membrane \cite{hu,lackner}. For length measurement purposes, 
this machinery is adiabatically equivalent to Eq. (\ref{wave}), meaning 
that it can be slowly deformed into scalar waves trapped in the 
bacterial body without changing its functionality.  The corresponding 
wave speed $v$ is about 0.15 $\mu$m/sec, the same as slow calcium waves.

The wave principle also has potential bearing on the overall shape of 
bacteria, most of which are cylinders of fixed width. The reasons for 
preferring this shape are not presently known. \cite{young} To measure 
the width of the body with a wave, one must excite the first azimuthal 
eigenmode of Eq. (\ref{wave}), also shown in Fig. \ref{f3}. This 
corresponds to a motion around the perimeter perpendicular to the body 
axis. This mode is necessarily doubly degenerate so long as the body is 
exactly cylindrical, so the corresponding form would not automatically 
be a cylinder unless the symmetry is broken, meaning that either 
right-handed or left-handed motion is preferred. This is a different 
issue from the handed spiral structures reported in bacterial walls 
\cite{huang1} because it requires also breaking of time-reversal 
symmetry, as occurs in a magnet.  Such symmetry breaking is known to 
occur in {\it E. coli}, where it manifests itself through preferred 
swimming handedness on glass slides, an effect attributable to a 
preferred rotation direction of the flagellum. \cite{diluzio,lauga}

The shape-regulating protein MreB has recently been observed to execute 
motion circumferential and perpendicular to the body axis in {\it B. 
subtilis} \cite{garner,dominguez,vanteefelen}. The experiments employ 
difficult sub-wavelength optical microscopy, and some details remain 
controversial. The reported azimuthal velocities range from 7 nm/sec to 
50 nm/sec, and one group reports a handedness bias while another rules 
it out. However, there is general agreement that MreB aggregates into 
small patches and that these translocate along the cell wall in a 
direction accurately perpendicular to the body axis.

In the context of mensuration it is not important whether the patch 
motion involves cell wall synthesis, as the experiments seem to suggest 
it does.  The principles by which diffusion is converted to wave motion 
are so general that they apply equally well to polymerization.

\section{Eucaryotes: Spindles and Syncytia}

Anything one says about eucaryotic size and shape control is necessarily 
speculative because so little definitive is known about it. However, the 
physical principles of resonant trapping are so simple and general that 
one might reasonably guess that they apply also to eucaryotic cells.

\begin{figure} 
\centerline{\includegraphics[width=.45\textwidth]{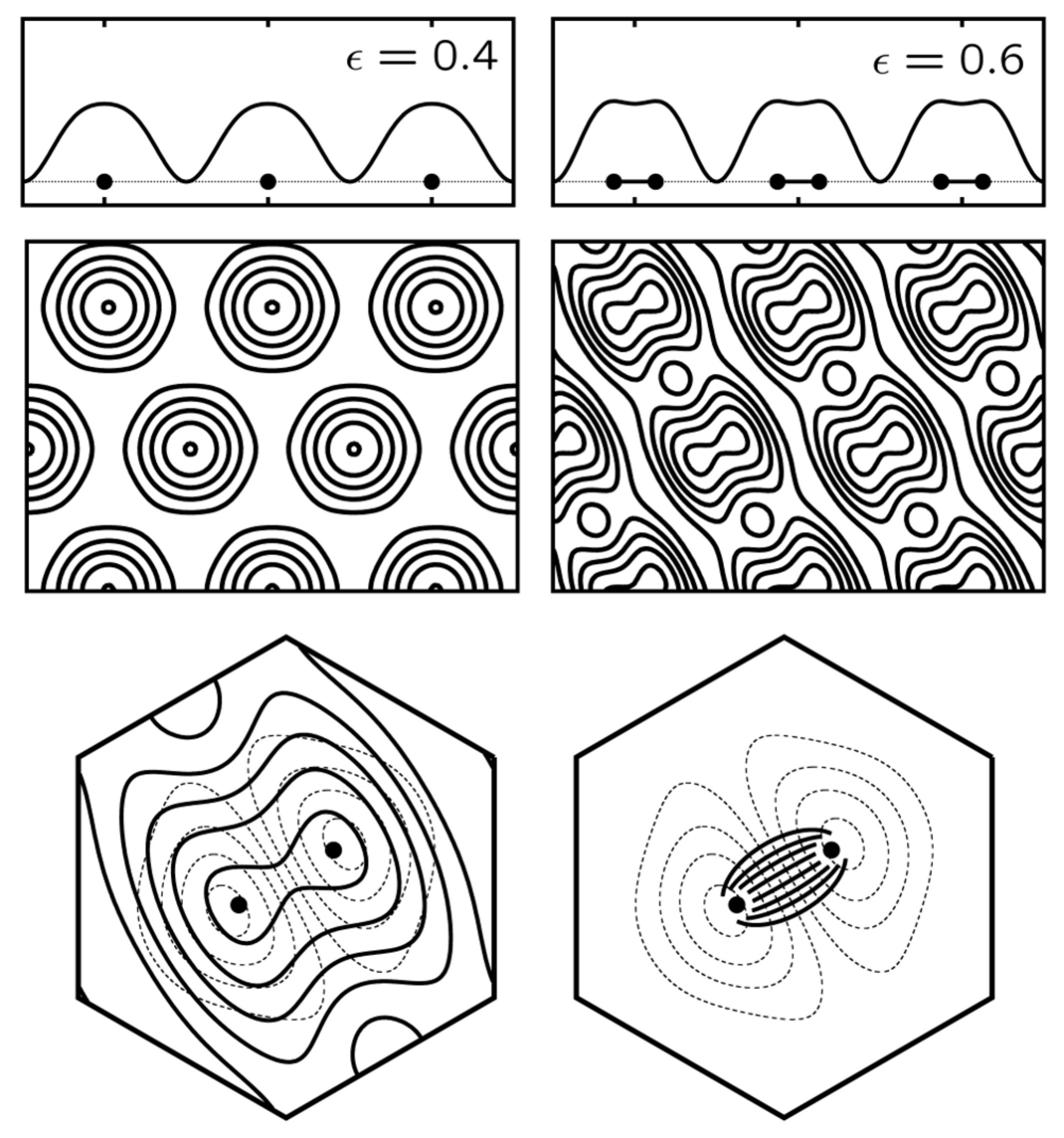}}
\caption{Illustration of standing wave syncytium.  Top: Plot of the time 
average of $|\psi(x,t)|^2 = | \cos( \omega x / v) \, \cos( \omega t )
+ \epsilon \sin( 2 \omega x / v ) \, \cos( 2 \omega t )
|^2$ at fixed frequency $\omega$ and wave speed $v$ for two different 
values of $\epsilon$. This shows how a standing wave grating
can be made to split centrosome locations by increasing the amplitude of 
its first harmonic.  Middle: The same quantity for the two-dimensional 
case expressed as a contour plot. The wavefunction is a ``twinkling 
eyes'' superposition of three plane waves at frequency $\omega$ and 
three more at $2 \omega$. \cite{yang} Bottom: Contour in the hexagonal 
unit cell showing phase-locked signal contours of $\psi$ and spindles 
constructed perpendicular to them, as in Fig. \ref{f3}.}
\label{f5} 
\end{figure}

Fig. \ref{f3} also shows the solution of Eq. (\ref{helmholtz}) for a 
spherical cavity, as might be appropriate for a eucaryotic cell. 
Everything is the same as for the pill-shaped cavity except for the 
boundary conditions, which we force to be mixed, thus pushing the 
antinode from the cell surface into the interior.  Mixed boundary 
conditions are appropriate for amplification machinery that resides 
partly in bulk interior and partly on or near the membrane.  Fig. 
\ref{f3} also shows trajectories generated by the rule of everywhere 
going downhill in the wavefunction gradient.  The similarity to the 
mitotic spindle is unmistakable.  Thus were an oscillating chemical 
potential field providing the navigation instructions for microtuble 
assembly, it would account quantitatively for (1) the location of 
centrosomes (the antinodes), (2) the existence and location of the 
metaphase plate, (3) the choice of a particular orientation for this 
plate, (4) the initiation and termination microtubules at the 
centrosomes, (5) their intersection at right angles with the metaphase 
plate, (6) their outward bulging at the plate, (7) the oblique angle 
formed between backward-going microtubules and the cell membrane, and 
(8) the observed scaling of the spindle assembly with cell size 
\cite{hazel,good}.

Waves also have the potential to account for the organization of 
structures without cell membranes. Fig. \ref{f5} shows a simple model of 
a syncytium made with standing waves. This specific construction uses 
three waves oriented at a physical angle $2\pi/3$ with respect to each 
other and also oscillating $2 \pi/3$ out of phase with each other in 
time so as to create a ``twinkling eyes'' dynamical pattern. \cite{yang} 
The recipe for locating the spindles is the same as in Fig. \ref{f3}.  
The equations used to generate Fig. \ref{f5} are much too primitive to 
describe an actual syncytium, among other reasons because the nuclei in 
real syncytia are not (cannot be) hexagonally arranged and because they 
have cytoskeletal structures where the cell membranes would normally 
have been.  Nonetheless Fig. \ref{f5} shows how standing waves can 
create organizational patterns beyond the immediate neighborhood of a 
specific nucleus, and thus how they might organize larger multi-cellular 
eucaryotic structures. There are some additional potential benefits, 
such as providing a natural signal to synchronize mitotic division and 
automatically scaling structures in a syncytial embryo to egg size.

\section{Conclusion: Static Versus Dynamic Reaction-Diffusion}

Dynamic reaction-diffusion is not a novel length mensuration method so 
much as an engineering advance over an older, more primitive one.  When 
implemented at the molecular level, it uses exactly the same chemistry 
that static reaction-diffusion does but simply manages time differently.  
An apt analogy would be the time management that distinguishes the 
Internet from the telegraph. Both use electricity to work, but the 
latter uses it more cleverly and is thus vastly more powerful. Dynamic 
mensuration is thus fully compatible with experimental evidence that 
small organisms use the static version often if the latter is imagined 
to be vestigial. \cite{meinhardt0,kondo,umulis}

The crucial engineering advantage of dynamic mensuration over the static 
variety is plasticity.  To measure out a length with an elementary 
diffusive morphogen one must balance a uniform destruction rate against 
a diffusion constant, a strategy that works perfectly well so long as 
the design is fixed. But if one wishes to change the design, one must 
modify the diffusion constant, the destruction rate, or both, making 
sure that the latter remains uniform.  If, on the other hand, one tunes 
the chemical reactions to manufacture waves with a fixed speed, lengths 
can be easily adjusted up or down simply by changing frequencies of 
stimulating oscillators. These need not be located in any particular 
place, for standing waves are rigid and thus insensitive to the location 
of their stimulus, an effect familiar from the operation of musical 
instruments. Thus the difficult hardware design need only be done once. 
The hardware can then be used again and again to measure out lengths of 
any size one likes, even with the latter changing on the fly in response 
to external events not encrypted in the genes.

The other important advantage of dynamic mensuration is 
generalizability.  Once the concept of turning diffusion into waves 
using amplifiers is discovered, it is very easy to imagine going by 
small steps to the invention of a sophisticated organ like the cochlea, 
which uses the same engineering principles but exploits mechanical 
diffusion, not chemical diffusion.  It is similarly easy to imagine 
going by small steps to the invention of neurons, which involve the same 
circuitry but with the amplifier gain turned up to make the propagating 
pulse nonlinear, and in which the underlying diffusive motion is 
electrical, not chemical. Diffusion is a very general physical 
phenomenon that results when when motion becomes disorganized.  The 
trick of reversing the descent into diffusive chaos using engines thus 
has applicability far beyond basic chemistry.

It is not a great concern that direct biochemical evidence for dynamic 
length measurement is thin. Bacteria, in particular, are very ancient, 
and it perfectly reasonable that they should employ both old and new 
technologies to form their bodies. But the more insightful observation 
is that laboratory detection of chemical potential oscillations is 
difficult and requires significant signal strength and integration time 
to do. Oscillations in other organisms might simply have not been found 
yet or be too weak or rapid to see easily.  The diffusion constants of 
MinD and MinE have been measured {\it in vivo} by fluorescence 
correlation spectroscopy to be roughly $D = 10 \mu{\rm m}^2$/sec 
\cite{meacci}. An amplifier time of $\tau = 10^{-3}$ sec, a number 
characteristic an ion channel protein, gives a maximum propagation 
velocity of $v = (D/\tau)^{1/2} = 100$ $\mu$m/sec, the speed of a fast 
calcium wave.  For a bacterium $3 \mu$m long, this gives a round-trip 
transit time of 0.06 sec.  Even faster speeds are possible with 
electrolyte ions, for which (ambipolar) diffusion constants are D 
$\cong$ 1000 $\mu$m$^2$/sec.

\begin{acknowledgments}
This work was supported by the National Science Foundation under
Grant No. PHY-1338376.
\end{acknowledgments}


\begin{thebibliography}{99}

\bibitem{nusslein} C. N\"{u}sslein-Volhard, ``The Identification of
        Genes Controlling Development in Flies and Fishes (Nobel
        Lecture),'' Angew. Chem. Int. Ed. Engl. {\bf 35}, 2176
        (1996).

\bibitem{stathopoulos} A. Stathopoulos and D. Iber, ``Studies of
        Morphogens: Keep Calm and Carry On,'' Development
        {\bf 140}, 4119 (2013).

\bibitem{sick} S. Sick, S. Reinker, J. Timmer, and T. Schlake, ``WNT and
        DKK Determine Hair Follicle Spacing Through a Reaction-Diffusion
        Mechanism,'' Science {\bf 314}, 1447 (2006).

\bibitem{economou} A. Economou {\it et al.}, ``Periodic Stripe Formation 
        by a Turing Mechanism Operating at Growth Zones in the Mammalian
        Palate,'' Nat. Gen. {\bf 44}, 348 (2011).

\bibitem{sheth} R. Sheth {\it et al.}, ``{\it Hox} Genes Regulate Digit
        Patterning By Cntrolling the Wavelength of a Turing-Type
        Mechanism,'' Science {\bf 338}, 1476 (2012).

\bibitem{muller} P. M\"{u}ller {\it et al.}, ``Differential Diffusivity
        of Nodal and Lefty Underlies a Reaction-Diffusion Patterning
        System,'' Science {\bf 336}, 721 (2012).

\bibitem{gregor} T. Gregor, A. P.  McGregor, and E. F. Wieschaus,
        ``Shape and Function of the Bicoid Morphogen Gradient in 
        Dipteran Species With Different Sized Embryos,''
        Devel. Biol. {\bf 316}, 350 (2008).

\bibitem{spirov} A. Spirov {\it et al.}, ``Formation of the Bicoid
        Morphogen Gradient: An mRNA Gradient Dictates the Protein
        Gradient,'' Development {\bf 136}, 605 (2009).

\bibitem{lipshitz} H. D. Lipshitz, ``Follow the mRNA: a New Model For
        Bicoid Gradient Formation,'' Nat. Rev. Mol. Cell Bio.
        {\bf 10}, 509 (2009).

\bibitem{grimm} O. Grimm, M. Coppey, and E. Wieschaus, ``Modeling the
        Bicoid Gradient,'' Development {\bf 137} 2253 (2010).

\bibitem{cheung} D. Cheung, C. Miles, M. Kreitman, and J. Ma,
        ``Adaptation of the Length Scale and Amplitude of the Bicoid 
        Gradient Profile to Achieve Robust Patterning in Abnormally
        Large {\it Drosophila melanogaster} Embryos,'' Development
        {\bf 141}, 124 (2014).

\bibitem{koch} A. J. Koch and H.  Meinhardt, ``Biological Pattern
        Formation: From Basic Mechanisms to Complex Structures,''
        Rev. Mod. Phys. {\bf 66}, 1481 (1994).

\bibitem{pearson} J. E. Pearson, ``Complex Patterns in a Simple 
        System,'' Science {\bf 261}, 189 (1993).

\bibitem{marshall} W. F. Marshall {\it et al.}, ``What Determines
        Cell Size?'' BMC Biol. {\bf 10}, 101 (2012).

\bibitem{nijhout} H. F. Nijhout, ``The Control of Body Size in
         Insects,'' Devel. Biol. {\bf 261}, 1 (2013).

\bibitem{stanger} B. Z. Stanger, ``Organ Size Determination and the
        Limits of Regulation,'' Cell Cycle {\bf 7}, 318 (2008).

\bibitem{king} R. S. King and P. A. Newmark, ``The Cell Biology of
        Regeneration,'' J. Cell. Biol. {\bf 196}, 553 (2012).

\bibitem{thompson} D. W. Thompson, (1992) {\it On Growth and Form: The
        Complete and Revised Edition} (Dover, New York, 1992).

\bibitem{turing} A. M. Turing, ``The Chemical Basis of Morphogenesis,''
        P. R. Soc. London B {\bf 237}, 37 (1952).

\bibitem{gold} T. Gold, ``Hearing. II. The Physical Basis of the
        Action of the Cochlea,'' P. R. Soc. Lond B {\bf 135},
        492 (1948).

\bibitem{camalet} S. Camalet, T. Duke, F. J\"{u}licher, and
        J. I. Prost, ``Auditory Sensitivity Provided By Self-Tuned
        Critical Oscillations of Hair Cells,'' Proc. Natl. Acad.
        Sci USA {\bf 97}, 3183 (2000).

\bibitem{hudspeth} A. J. Hudspeth, F. J\"{u}licher, and P. Martin,
        ``A Critique of the Critical Cochlea: Hopf - A Bifurcation 
        - Is Better Than None,'' J. Neurophysiol. {\bf 104}, 1219 
        (2010).

\bibitem{scott} A. W. Scott, ``The Electrophysics of a Nerve Fiber,''
        Rev. Mod. Phys. {\bf 47}, 487 (1975).

\bibitem{goldbeter} A. Goldbeter, ``Oscillations and Waves of Cyclic
        AMP in {\it Dictyostelium}: A Prototype For Spatio-Temporal
        Organization and Pulsatile Intercellular Communication,''
        Bull. Math. Biol. {\bf 58}, 1095 (2006).

\bibitem{jaffe} L. F. Jaffe, ``Calcium waves,'' Phil. Trans.
        Roy. Soc. B {\bf 363}, 1311 (2008).

\bibitem{jaffe1} L. F. Jaffe, ``Organization of Early Development By
        Calcium Patterns,'' Bioessays {\bf 21}, 567 (1999).

\bibitem{whitaker} M. Whitaker and J. Smith, ``Introduction. Calcium
        Signals and Developmental Patterning,'' Phil. Trans. Roy.
        Soc. B, {\bf 363}, 1307 (2008).

\bibitem{weissman} T. A. Weissman, P. A. Riqueime, L. Ivic,
        A. C. Flint, and A. R. Kriegstein, ``Calcium Waves Propagate 
        Through Radial Glial Cells and Modulate Proliferation in the
        Developing Neocortex,m'' Neuron {\bf 43}, 647 (2004).
 
\bibitem{scemes} E. Scemes and C. Giaume, ``Astrocyte Calcium Waves:
        What They Are and What They Do,'' Glia {\bf 54}, 716 (2006).

\bibitem{kuga} N. Kuga, T. Sasaki, Y. Takahara, N. Matsuki, and Y.
        Ikegaya, ``Large-Scale Calcium Waves Traveling Through 
        Astrocytic Networks {\it in vivo},'' J. Neurosci. {\bf 31},
        2607 (2011).

\bibitem{junkin} M. Junkin, Y. Lu, J. Long, P. A. Deymier, J. B. Hoying, 
        and P. K. Wong, ``Mechanically Induced Intercellular Calcium
        Communication in Confined Endothelial Structures,'' Biomat.
        {\bf 34}, 2049 (2013).

\bibitem{choi} W.-G. Choi, M. Toyota, S.-H. Kim, R. Hilleary, and
        S. Gilroy, ``Salt Stress-Induced Ca$^{2+}$ Waves Are Associated
        With Rapid, Long-Distance Root-to-Root Signaling in Plants,''
        Proc. Natl. Acad. Sci. {\bf 111}, 5497 (2014).     

\bibitem{lenz} P. Lenz and L. Sogaard-Anderson, ``Temporal and Spatial
        Oscillations in Bacteria,'' Nat. Rev. Microbiol. {\bf 9},
        565 (2011).

\bibitem{siegman} A. E. Siegman, {\em Lasers} (University Science
        Books, Herdon, VA, 1986)

\bibitem{gale} J. E. Gale and J. F. Ashmore, ``An Intrinsic Frequency
        Limit to the Cochlear Amplifier,'' Nature {\bf 389}, 63 (1997).

\bibitem{ashmore} J. Ashmore {\it et al.}, ``The Remarkable Cochlear 
        Amplifier,'' Hearing Res. {\bf 266}, 1 (2010).

\bibitem{manley} G. A. Manley, ``Evidence For an Active Process and
        a Cochlear Amplifier in Nonmammals,'' J. Neurophysiol.
        {\bf 86}, 541 (2001).

\bibitem{warren} B. Warren, A. N. Lukashkin, and I. J. Russell, ``The 
        Dynein-Tubulin Motor Powers Active Oscillations and Amplification
        in the Hearing Organ of the Mosquito,'' P. R. Soc. B
        {\bf 277}, 1761 (2010).

\bibitem{mhatre} N. Mhatre and D. Robert, ``A Tympanal Insect Ear 
        Exploits a Critical Oscillator For Active Amplification and
        Tuning,'' Curr. Biol. {\bf 23}, 1952 (2013).

\bibitem{mora} E. C. Mora, A. Cobo-Cuan, F. Mac\'{i}as-Escriv\'{a},
        M. P\'{e}rez, M. Nowotny, and M. K\"{o}ssl, ``Mechanical Tuning
        of the Moth Ear: Distortion-Product Otoacoustic Emissions
        and Tympanal Vibrations,'' J. Exp. Biol. {\bf 216}, 3863 (2013). 

\bibitem{hodgkin} A. L. Hodgkin and A. F. Huxley, ``A Quantitative
        Description of Membrane Current and Its Application to
        Conduction and Excitation of Nerve,'' J. Physiol.
        {\bf 117}, 500 (1952).

\bibitem{fitzhugh} R. Fitzhugh, ``Mathematical Models of Excitation
        and Propagation in Nerve,'' in {\it Biological Engineering}, ed.
        by H. P. Schwan (McGraw Hill, New York, 1969).

\bibitem{goldin} A. L. Goldin, ``Mechanisms of Sodium Channel 
        Inactivation,'' Curr. Opin. Neurobiol. {\bf 13}, 284 (2003).

\bibitem{desurvire} E. Desurvire, ``Study of the Complex Atomic
        Susceptibility of Erbium-Doped Fiber Amplifiers,'' J.
        Lightwave Technol. {\bf 8}, 1517 (1990).

\bibitem{toll} J. S. Toll, ``Causality and the Dispersion Relation:
        Logical Foundations,''  Phys. Rev. {\bf 104}, 1760 (1956).

\bibitem{showalter} K. Showalter and J. J. Tyson, ``Luther's 1906
        Discovery of Chemical Waves,'' J. Chem. Ed. {\bf 64}, 742
        (1987).

\bibitem{kemp} D. T. Kemp, ``Stimulated Acoustic Emissions From Within
        the Human Auditory System,'' J. Acoust. Soc. Am.  {\bf 54}, 1386 
        (1978).

\bibitem{pelling} A. E. Pelling, S. Dehati, E. B. Gralla, J. B. 
        Valentine, and J. K. Gimzewski, ``Local Nanomechanical Motion of 
        the Cell Wall of {\it Saccharomyes cerevisiae},'' Science
        {\bf 305}, 1147 (2004).

\bibitem{raskin} D. M. Raskin and P. A. J. de Boer, ``Rapid Pole-to-Pole
        Oscillations of a Protein Required For Directing Division
        to the Middle of {\it Escherichia coli},'' Proc. Natl. Acad. Sci.
        USA {\bf 96}, 4971 (1999).

\bibitem{meacci-thesis} G. Meacci, {\it Min Oscillations in Escherichia
        Coli} (VDM, Saarbr\"{u}ken, 2009).

\bibitem{juarez} J. R. Juarez and W. Margolin, ``Changes in the Min 
        Oscillation Pattern Before and After Cell Birth,'' J.
        Bacteriol. {\bf 192}, 4134 (2010).

\bibitem{marston} A. L. Marston, H. B. Thomaides, D. H. Edwards, M. E.
        Sharpe, and J. Errington, Polar Localization of the MinD Protein
        of {\it Bacillus subtilis} and Its Role in Selection of the 
        Mid-Cell Division Site,'' Genes Devel. {\bf 12}, 3419 (1998().

\bibitem{howard} M. Howard, A. D. Rutenberg, and S. De Vet, ``Dynamic
        Compartmentalization of Bacteria: Accurate Division in
        {\it E. coli},'' Phys. Rev. Lett. {\bf 87}, 278102 (2001).

\bibitem{meinhardt5} H. Meinhardt and P. A. J. de Boer, ``Pattern 
        Formation in {\it Escherichia coli}: A Model For the Pole-to-Pole
        Oscillations of Min Proteins and the Localization of the
        Division Site,'' Proc. Natl. Acad. Sci. USA {\bf 98}, 14202
        (2001).

\bibitem{huang} K. C. Huang, Y. Meir, and N. S. Wingreen, ``Dynamic 
        Structures in {\it Escherichia coli}: Spontaneous Formation of 
        MinE Rings and MinD Polar Zones,'' Proc. Natl. Acad. Sci. (USA)
        {\bf 100}, 12724 (2003).

\bibitem{bonny} M. Bonny, E. Fischer-Friedrich, M. Loose, P. Schwille,
        and K. Kruse, ``Membrane Binding of MinE Allows For a 
        Comprehensive Description of Min-Protein Pattern Formation,''
        PLOS Comp. Biol. {\bf 9}, e1003347 (2013).

\bibitem{halatek} J. Halatek and E. Frey, ``Effective 2d Model Does Not
        Account For Geometry Sensing By Self-Organized Proteins 
        Patterns,'' Proc. Natl. Acad. Sci. USA {\bf 111}, E1817 (2014).

\bibitem{adler} H. I. Adler, W. D. Fisher, A. Cohen, and A. A. 
        Hardigree, ``Miniature {\it Escherichia coli} Cells Deficient in 
        DNA,'' Proc. Natl. Acad. Sci. (USA) {\bf 57}, 321 (1967).

\bibitem{deboer} P. A. J. de Boer, R. E. Crossley, and L. I. Rothfield,
        ``A Division Inhibitor and a Topological Specificity Factor Coded 
        For by the Minicell Locus Determine Proper Placement of the 
        Division Septum in {\it E. coli},'' Cell {\bf 56}, 641 (1989).

\bibitem{loose} M. Loose, E. Fischer-Friedrich, J. Ries, K. Kruse, and
        P. Schwille, ``Spatial Regulators For Bacterial Cell Division
        Self-Organize into Surface Saves {\it in Vitro},'' Science
        {\bf 320}, 789 (2008).

\bibitem{ivanov} V. Ivanov and K. Mizuuchi, ``Multiple Modes of
        Interconverting Dynamic Pattern Formation by Bacterial Cell
        Division Proteins,'' Proc. Natl. Acad. Sci. (USA) {\bf 107},
        8071 (2010).

\bibitem{schweizer} J. Schweizer, M. Loose, M. Bonny, K. Kruse, 
        I M\"{o}nch, and P. Schwille, ``Geometry Sensing By 
        Self-Organized Protein Patterns,'' Proc. Natl. Acad. Sci.
        (USA) {\bf 109}, 15283 (2012).

\bibitem{hu} Z. Hu, E. P. Gogol, and J. Lutkenhaus, Dynamic Assembly of
        MinD on Phospholipid Vesicles Regulated by ATP and
        MinE,'' Proc. Natl. Acad. Sci. (USA) {\bf 99}, 6761 (2002).

\bibitem{lackner} L. L. Lackner, D. M. Raskin, and P. A. J. de Boer,
        ``ATP-Dependent Interactions Between {\it Escherichia
        coli} Min Proteins and the Phospholipid Membrane P{\it in
        vitro},'' J. Bacteriol. {\bf 185}, 735 (2003).

\bibitem{young} K. D. Young, ``Bacterial Shape: Two-Dimensional 
        Questions and Possibilities,'' Annu. Rev. Microbiol. 
        {\bf 64}, 223 (2010).

\bibitem{huang1} K. C. Huang, D. W. Ehrhardt, and J. W. Shaevitz, ``The 
        Molecular Origins of Chiral Growth in Walled Cells,''
        Curr. Opin. Microbiol. {\bf 15}, 1 (2012).

\bibitem{diluzio} W. R. DiLuzio, L. Turner, M. Mayer, P. Garstecki,
        D. B. Weibel, H. C. Berg, and G. M. Whitesides, ``{\it Escherichi
        coli} Swim on the Right-Hand Side,'' Nature {\bf 435}, 1271 
        (2005).

\bibitem{lauga} E. Lauga, W. R. DiLuzio, G. M. Whitesides, and M. A.
        Stone, ``Swimming in Circles: Motion of Bacteria Near Solid 
        Boundaries,'' Biophys. J. {\bf 90}, 400 (2006).

\bibitem{garner} E. C. Garner, R. Bernhard, W. Wang, X. Zhuang, D. Z. 
        Rudne, and T Mitchison, ``Coupled, Circumferential Motions of the
        Cell Wall Synthesis Machinery and MreB Filaments in {\it B.
        subtilis},'' Science {\bf 333}, 222 (2011).

\bibitem{dominguez} J. Dom\'{i}inguez-Escobar, A. Chastanet, A. H.
         Crevenna, V. Fromion, R. Wedlich-S\"{o}ldner, and R 
         Carballido-L\'{o}pez, ``Precessive Movement of MreB-Associated 
         Complexes in Bacteria,'' Science {\bf 333}, 225 (2011).

\bibitem{vanteefelen} S. van Teeffelen, S. Wang, L. Furchgott, K. C. 
        Huang, N. S. Wingreen, J. W. Shaevitz, and Z. Gital, ``The 
        Bacterial Actin MreB Rotates, and Rotation Depends on Cell-Wall 
        Assembly,'' Proc. Natl. Acad. Sci. {\bf 108}, 15882 (2011).

\bibitem{hazel} J. Hazel {\it et al.}, ``Changes in Cytoplasmic Volume
        Are Sufficient to Drive Spindle Scaling,'' Science
        {\bf 342}, 853 (2013).

\bibitem{good} M. C. Good, M. D. Vahey, A. Skandarjah, D. A. Fletcher,
        and R. Heald, ``Cytoplasmic Volume Modulates Spindle Size During
        Embryogenesis,'' Science {\bf 324}, 856 (2013).

\bibitem{yang} L. Yang and I. R. Epstein, ``Oscillatory Turing Patterns in
        Reaction-Diffusion Systems With Two Coupled Layers,'' Phys.
        Rev. Lett. {\bf 90}, 178303 (2003).

\bibitem{meinhardt0} H. Meinhardt H, ``Models of Biological Pattern
        Formation: From Elementary Steps to the Organization of
        Embryonic Axes,''  Curr. Top. Dev. Biol. {\bf 81}, 1 (2008).

\bibitem{kondo} S. Kondo and T. Miura, ``Reaction-Diffusion Model as
        a Framework for Understanding Biological Pattern Formation,''
        Science {\bf 329}, 1616 (2010).

\bibitem{umulis} D. M. Umulis and H. G. Othmer, ``Mechanisms of
        Scaling in Pattern Formation,'' Development {\bf 140},
        4830 (2013).

\bibitem{meacci} G. Meacci, J. Ries, E. Fischer-Friedrich, N. Kahya,
        P. Schwille, and K. Kruse, ``Mobility of Min-proteins in
        {\it Escherichia coli} Measured by Fluorescence Correlation
        Spectroscopy,''  Phys. Biol. {\bf 3}, 255 (2006).

\end{thebibliography}
\end{document}